# Robustness of Majorana modes to potential disorder in Fe chains on a superconducting Rashba alloy

*Harim Jang[1,*], Daniel Crawford[2], Khai Ton That[1], Lucas Schneider[1], Jens Wiebe[1], Makoto Shimizu[3], Harald O. Jeschke[4], Stephan Rachel[5], and Roland Wiesendanger[1,*]*

[1]Department of Physics, University of Hamburg; D-20355 Hamburg, Germany.
[2]Department of Physics and Nanoscience Center, University of Jyväskylä; P.O. Box 35 (YFL), FI-40014 University of Jyväskylä, Finland.
[3]Department of Physics, Graduate School of Science, Kyoto University; Kyoto 606-8502, Japan.
[4]Research Inst. for Interdisciplinary Science, Okayama University; Okayama 700-8530, Japan.
[5]School of Physics, University of Melbourne; Parkville, Victoria 3010, Australia.

**\*Corresponding authors.**

**Email: harim.jang@uni-hamburg.de, roland.wiesendanger@uni-hamburg.de**

**Majorana modes offer great potential for fault-tolerant quantum computation due to their topological protection. However, for superconductor–semiconductor nanowire hybrids, intrinsic disorder makes the unambiguous detection of Majorana modes difficult. Here, we construct 1D spin chains from individual Fe atoms on the Rashba surface alloy BiAg$_2$/Ag(111) with proximity-induced superconductivity from a Nb(110) substrate. While the Fe chains exhibit perfect crystalline order, we observe nano-scale potential disorder of the BiAg$_2$/Ag(111)/Nb(110) heterostructure by scanning tunneling microscopy. However,**

**this does not prevent the emergence of zero-energy modes at both ends of the Fe chains, in agreement with tight-binding calculations showing that they are only found in the topologically non-trivial regime of the phase diagram. These Majorana modes are indeed robust against potential disorder.**

Atomic magnetic chains on superconducting (SC) substrates have been proposed as a promising platform for the observation of topological superconductivity and associated zero-energy Majorana modes (*1-8*). Signatures for Majorana modes have first been found experimentally for self-assembled Fe chains on a SC Pb(110) substrate by detecting zero-energy peaks in the differential tunneling conductance spectra measured by scanning tunneling spectroscopy (STS) at the ends of such Fe chains (*9-11*). However, a topological gap could not be detected for this particular magnet-superconductor hybrid (MSH) system even down to a temperature of 30 mK. Moreover, the influence of disorder, as present for these self-assembled Fe chains on Pb(110), could not be clarified at that time. Subsequent STS studies on disorder-free atom-by-atom constructed Fe chains on SC Re(0001) confirmed the emergence of zero-energy states at both ends of perfectly ordered chains as a function of chain length, while a topological gap could not be detected at the measurement temperature of 300 mK (*12, 13*). Further artificially designed MSH platforms, e.g. based on quantum spin chains (*14*) or SC alloy substrates (*15*) have been investigated. Recently, bottom-up fabricated Mn chains on ultimately clean SC Nb(110) substrates allowed the experimental observation of a spin-orbit-coupling induced gap as large as 180 $\mu$eV (*16*) in one of the multi-orbital Shiba bands and chain-length dependent oscillations of the low-energy modes simultaneously probed by STS at both ends of disorder-free atomic chains (*17*).

These hybridization-induced splitting oscillations have been theoretically identified early on as a smoking gun for the experimental confirmation of the elusive Majorana modes (*18, 19*).

Here, we present an exciting novel model-type MSH system consisting of atom-by-atom constructed Fe chains of various lengths on a $BiAg_2$/Ag(111) surface alloy with proximity-induced superconductivity from a Nb(110) substrate. The $BiAg_2$ surface alloy grown epitaxially on a single-crystalline Ag(111) thin film is known to exhibit a large Rashba-type spin-orbit coupling (*20-22*) which is expected to favour a large topological gap of the MSH system comprising a SC Nb(110) substrate, i.e., the elemental superconductor with the highest SC transition temperature of 9.3 K. We show that the long-range nature of the Yu-Shiba-Rusinov (YSR) states of Fe atoms residing on hollow-sites with respect to the $BiAg_2$ lattice allows for significant hybridization between the Fe atoms even for a spacing of two atomic lattice sites. Based on low-temperature STS measurements of the Fe chains on the SC $BiAg_2$/Ag(111)/Nb(110) substrate as a function of chain length, the emergence of highly localized zero-energy modes at both ends of these perfect crystalline Fe chains can directly be observed in real space. Tight-binding calculations suggest that this MSH system resides in a topologically non-trivial regime and that the experimentally observed zero-energy end states can be associated with Majorana modes. Interestingly, the $BiAg_2$/Ag(111)/Nb(110) substrate is found to exhibit potential disorder, in contrast to earlier experiments involving SC single-crystalline Re(0001) and Nb(110) substrates. The potential disorder of the $BiAg_2$/Ag(111)/Nb(110) substrate, as revealed by atomic-resolution STM measurements, affects the spatial distribution of the finite-energy YSR bulk states, but does not prevent the observation of the zero-energy end states, in agreement with model calculations taking a similar potential disorder distribution as observed experimentally into account. Our results provide direct proof for the robustness of Majorana modes in MSH systems even in the presence of disorder (*23, 24*) and can additionally explain earlier observations of zero-energy end states in disordered Fe chains on SC Pb(110) substrates (*10*).

**Site-dependent YSR states of single Fe atoms and their spatial extension on a superconducting BiAg$_2$ surface alloy**

In order to combine large spin-orbit coupling with s-wave superconductivity, as an important ingredient for achieving a topological SC state, we have designed and realized a novel type of heterostructure consisting of a SC Nb(110) substrate, epitaxially grown Ag(111) islands with proximity-induced superconductivity (*25, 26*), and a monolayer of a BiAg$_2$(111) surface alloy on top exhibiting a √3×√3*R*30° superstructure (see Figs. 1A,D). Differential tunneling conductance (d*I*/d*V*) measurements performed at $T = 4.2$ K on top of the BiAg$_2$ surface alloy clearly reveal the characteristics of an induced SC state with an energy gap $\Delta_S$ of 1.31 meV (see Fig. 1G). Since a SC Nb-tip was used to enhance the energy resolution, the Fermi level ($E_F$) of the sample is shifted by the size of the SC energy gap $\Delta_T$ of the Nb-tip, as indicated by a dashed line in Fig. 1G. A numerically deconvoluted spectrum is presented in Fig. 1J (see Suppl. Mater., section 1, for details of the deconvolution procedure). Individual Fe atoms were subsequently deposited on the cold BiAg$_2$/Ag(111)/Nb(110) substrate, thereby preventing surface diffusion. Atomic-resolution STM measurements reveal two distinct adsorption sites for single Fe adatoms, as shown in Figs. 1E,F, namely 'bridge' and 'hollow' sites, for which the Fe adatoms are coordinated by two or three Bi atoms, respectively (see Figs. 1B,C).

The d*I*/d*V* spectra measured above individual Fe adatoms on the SC BiAg$_2$/Ag(111)/Nb(110) substrate exhibit discrete bound states inside the SC gap, so-called YSR states (see Figs. 1H,I for the as-measured spectra and Figs. 1K,L for the deconvoluted spectra). For the Fe bridge-site adatom, there are clearly visible peaks at $E-E_F = \pm 0.40$ meV related to a pair of YSR in-gap states, while the spectroscopic features at $E-E_F = \pm 1.3$ meV are caused by the SC coherence peaks. In contrast, the tunneling spectrum of the Fe hollow-site adatom exhibits a pronounced YSR state located at $E_F$.

Besides the strong adsorption-site dependence of the YSR bound state energies, there is also a significant difference in the spatial extension of YSR states for Fe adatoms on bridge- and hollow-sites. Figure 1M shows a large-scale constant-current STM image of the BiAg$_2$/Ag(111)/Nb(110) surface with several Fe adatoms adsorbed on either bridge- or hollow-sites (examples are indicated by a dashed green arrow and a yellow arrow, respectively). Interestingly, the corresponding spectroscopic d$I$/d$V$ map obtained for $eV = -E_F$ (Fig. 1N) reveals that the pronounced YSR state of hollow-site Fe adatoms exhibits long-range oscillations over more than ten nanometers with a periodicity of around 2 nm (see yellow arrow in Fig. 1N), while the bridge-site Fe adatoms show states spatially localized to atomic-scale dimensions (indicated by the green arrow in Fig. 1N). The amplitude of the long-range YSR state oscillations decays inversely proportional to the distance from the hollow-site Fe adatoms, suggesting that the YSR bound states originate from an exchange coupling to an effectively 2D superconductor associated with the BiAg$_2$ monolayer film (see Suppl. Mater., section 2).

Based on our atomic-scale spectroscopic studies of the site-specific YSR states, we can conclude that the hollow-site Fe adatoms are most promising as elemental building blocks for the construction of spin chains on BiAg$_2$/Ag(111)/Nb(110) in view of achieving a topologically non-trivial SC state: first, the YSR state energy is already close to $E_F$, making it a promising candidate for engineering YSR bands crossing $E_F$. Second, the YSR state of the hollow-site Fe adatoms is of long-range nature, thereby facilitating the hybridization between YSR states of Fe adatoms even at nm-scale distances and therefore the formation of YSR bands.

**Atom-by-atom construction of ferromagnetic Fe chains on a superconducting BiAg$_2$ Rashba alloy and the emergence of zero-energy edge states**

We crafted well defined spin chains by individual Fe-atom manipulation to hollow-sites, linearly arranged along the [1$\bar{1}$0]-direction with respect to the BiAg$_2$(111) lattice, and with a given interatomic spacing of $2a \sim 0.97$ nm. Topographic STM snapshots of the systematic construction of Fe chains with a variable length of two (Fe-2) up to eleven (Fe-11) atoms are shown in Fig. 2A. By using spin-polarized STM, we confirmed a ferromagnetically ordered state of such atomic Fe chains, in agreement with ab initio calculations (see Suppl. Mater., sections 3 and 8).

Energy-dependent d$I$/d$V$ line profiles were obtained along all Fe-$n$ chains, and some representative as-measured as well as numerically deconvoluted data sets are presented in Figs. 2C-E and Figs. 2F-H, respectively (see Suppl. Mater., section 4, for the evolution of the spectroscopic d$I$/d$V$ line profiles for all Fe-$n$ chains). For Fe-$n$ chains with $n > 2$, pronounced tunneling conductance peaks at $E_F$ (as measured with the SC Nb-tip), i.e., zero-energy edge states in the deconvoluted spectra, were found at both ends of the chains. It is worth noting that the spatially highly localized zero-energy edge states emerging in Fe-$n$ chains ($n > 2$) have a fundamentally different origin compared to the YSR state of an individual hollow-site Fe adatom being located at $E_F$, because the sizable hybridization between YSR states of neighboring Fe adatoms lifts the degeneracy of the YSR bound state energies for Fe-2 in the $2a$- and even for $3a$-chains (see Suppl. Mater., section 5). Inside the chains, the YSR band develops at around ±0.35 meV, resulting from the hybridization between the YSR states of neighboring Fe hollow-site adatoms.

Fig. 2I presents spectroscopic d$I$/d$V$ maps of the Fe-11 chain at energies of $\pm E_F$ and $\pm(E_F + 0.35$ meV). Zero-energy edge states clearly show up by significant spectral weight only for the Fe hollow-site adatoms at the chain ends and only at $E_F$. On the other hand, the d$I$/d$V$ intensity at both ends is significantly suppressed at finite energies of the YSR bands, which only show up inside the chain. A direct comparison between as-measured tunneling spectra of edge (Fe(1)) and middle atoms (Fe(3)) of the Fe-11 chain is presented in Fig. 2J (these positions are highlighted by vertical dashed lines in Fig. 2I).

**Intrinsic disorder in the BiAg$_2$ surface potential**

When examining the spectroscopic d$I$/d$V$ maps of the Fe-11 chain at finite energies (see, e.g., Fig. 2I top and bottom) in more detail, one can recognize a spatial variation of the d$I$/d$V$ intensity from one Fe atom to the next. This is associated with a local variation of the YSR state energies and contributes to the observed width of the YSR band (Figs. 2F-H). We have carefully investigated the origin of this effect and found that it is connected with a nanoscale surface potential disorder typical for the BiAg$_2$ surface alloy prepared on Ag(111)/Nb(110) and not found on bare surfaces of elemental superconductors, such as Re(0001) (*11, 12*) or Nb(110) (*13,14*). Even though the Bi atoms form a well ordered hexagonal lattice (see Fig. 1D), a spatial variation of the local density of states (LDOS) on the nanoscale can clearly be recognized in large-scale constant-current STM images (see Fig. 2K). The observed nanoscale electronic inhomogeneity most likely originates from local strain effects in the bulk of Ag(111) nanostructures grown on Nb(110) (*27*) or from structural inhomogeneities at the Ag(111)/Nb(110) interface . The degree of disorder can be quantified by the distribution of recorded relative Z-height values above the Bi-atom sites, where stronger disorder increases the width of the distribution (see Fig. 2K, right). Despite the presence of intrinsic potential disorder on the BiAg$_2$ surface, which affects the spatial distribution of the finite-energy YSR bulk states, it does not influence the spatial homogeneity of the proximity-induced superconducting state of the BiAg$_2$ surface alloy (see Suppl. Mater., section 6) and it does not prevent the observation of the zero-energy end states of the atomic Fe chains on the SC BiAg$_2$/Ag(111)/Nb(110) heterostructure (Fig. 2I).

**Robustness of Majorana zero modes against potential disorder in magnet-superconductor hybrids**

To support the interpretation of the experimental results, we construct a minimal tight-binding model involving superconductivity to analyze atomic Fe chains on BiAg$_2$/Ag(111)/Nb(110) (see Suppl. Mater., section 9). To this end, we only consider the top surface Bi atoms forming a hexagonal lattice with period $a$ as the substrate, where Fe atoms are placed on the hollow sites of the Bi lattice with distance $2a$ between neighboring Fe atoms (see Fig. 3A). The Bi atoms are coupled to an isotropic SC $s$-wave order parameter $\Delta_0$, while the Fe atoms have a magnetic moment of size $J$. Thus, our model mimics the SC proximity effect, when we observe (topological) superconductivity on the Fe chain. Despite being minimal, the model's parameter space is far too large; our strategy here is to use experimental insights and results from ab initio calculations, to reduce this parameter space as much as possible and in the most realistic way. For the normal state, we fit a parabolic, spin-split substrate band with standard Rashba spin-orbit coupling to the quasiparticle interference measurements of Ref. (*28*) (see Fig. 3B and Suppl. Mater., section 9). This allows us to determine reasonable parameters for the hopping, chemical potential, and Rashba spin-orbit coupling $(t, \mu, \alpha) = (100, -545, 25)$ meV of the substrate, by matching the Rashba energy $E_R$ and scattering vector $q$. While $\Delta_{Nb} = 1.51$ meV, and up to 87% of that was measured on the BiAg$_2$ surface thanks to a strong SC proximity effect, we must use for our purely 2D model a significantly larger value of $\Delta_0$, to account for the absence of a 3D SC substrate. While in experiment the proximity-induced superconducting gap measured on the chain $\Delta$ is similar to the bulk gap $\Delta_0$, in our model the absence of a 3D SC substrate results also in $\Delta < \Delta_0$. We choose $\Delta_0 = 70$ meV and a value for the Fe-Bi hybridization of $\Gamma = 35$ meV to match the experimentally observed spectral features in Fig. 2J, corresponding to an upper limit of the minigap of $\Delta_{mini}/\Delta \sim 0.05/1.5$, estimated from the experimental energy resolution, and a YSR bandwidth $\sim \Delta/4$ (see Figs. 2E,H). Next, we perform density functional theory (DFT) calculations (see Suppl. Mater., section 8) for $2a$-Fe chains on BiAg$_2$/Ag(111)/Nb(110), which turn out to be ferromagnetically coupled ($\approx -1.3$ meV for unit moments) with energetically preferred out-of-plane moments, in line with the SP-STM measurements (see Suppl. Mater., section 3). The remaining parameters are the chemical potential of the Fe atoms, $\mu_{Fe}$, and the effective magnetic moment $J$. In order to guarantee that the energy of the single YSR state of a hollow-site Fe adatom assumed in the model is $E_{YSR} \approx 0$, as observed experimentally (see Figs. 1I,L), we have to satisfy the relation $\mu \approx J + \Gamma/3$. The resulting one-dimensional model is in topological class D (*29, 30*) with a $Z_2$ invariant. This is readily verified by computing the $Z_2$ invariant for an infinitely long chain (periodic boundary conditions), leading to the phase diagram in Fig. 3C. Next, we focus on chain length $L = 11$ on a substrate with $400 \times 10$ Bi atoms and open boundary conditions. Such large substrates are necessary to observe the SC coherence peaks of the substrate at $E \approx \Delta_0$. In Fig. 3D we show a typical phase diagram depending on $J$ and $\mu_{Fe}$, featuring a sizeable region of topological superconductivity. The topologically non-trivial regime is identified by (i) states at $E = 0$ (indicated by the white dashed line in Fig. 3D) and (ii) a finite gap $E_1 - E_0$, shown as color plot in Fig. 3D. The cross (plus) in Fig. 3D corresponds to parameters guaranteeing topological (trivial) superconductivity and $E_{YSR} \approx 0$ of a single impurity on a hollow site.

We consider three cases: (i) an $L = 11$ chain with $J = 18$ meV (corresponding to the plus in Fig. 3D), being topologically trivial with an induced minigap $\Delta_{triv} = 2$ meV; (ii) an $L = 11$ chain with J = 22 meV (corresponding to the cross in Fig. 3D), being topologically non-trivial with an induced minigap $\Delta_{topo} = 2$ meV; (iii) we also show an example for a longer chain (L = 30 chain, J = 22 meV, topological regime, induced minigap $\Delta_{topo} = 2$ meV) to demonstrate that our results are robust upon varying the system size ($\mu = 30$meV for all three cases). We show the LDOS($E, x$) along the chain for these three cases in Fig. 3E-G, along with the corresponding LDOS($x, y$) for $E = 0$ close to the chain, and the LDOS($x, y$) at the energy of the band gap. In both the LDOS($E, x$) and LDOS($x,y$) at $E = 0$, localized Majorana modes are indicative of the topological phase and can be clearly distinguished from bulk states. In the trivial phase, there are no zero-energy end states.

We now turn to the disorder-analysis of the BiAg$_2$ surface alloy. We model the disordered BiAg$_2$ surface by applying a correlated potential disorder $\mu_i$ to the substrate sites, see Fig. 2L. We use a Gaussian noise source similar to the experimental data (Fig. 2K), and apply a low-pass filter (see Suppl. Mater., section 10) to eliminate short-wavelength correlations. We choose a cutoff frequency such that there are nanoscale correlations in the disorder distribution (Fig. 2L), similar to experiment (Fig. 2K). We have selected the strength D of the disorder distribution such that the hard minigap just closes and only a soft minigap remains; at the same time, the Majorana modes remain well-defined and are clearly visible within the LDOS(E, x) plots (see Figs. 4D,E). Furthermore, we choose the width of the Gaussian source $D$ such that the resulting LDOS is similar to experiment, up to $D \sim \Delta$. We compute 100 disorder realizations based on the three cases discussed above. We plot the disorder averaged DOS (see Figs. 4A-C) taken at one chain end (the DOS at the other end is essentially identical after disorder-averaging), highlighting the effects of disorder on the YSR band and the end states. While disorder can localize zero-energy states, in none of our simulations do we find spurious zero-energy end states in the trivial phase. In contrast, the zero-bias peak in the DOS due to the Majorana modes is robust to disorder over an extended range of disorder strengths. Due to the spin-triplet correlations, in all cases disorder results in a soft minigap (see Suppl. Mater., section 10). We also show the LDOS($E$, $x$) in the topological phase for a single but representative disorder realization (see Figs. 4D,E), for $L = 11$ and $L = 30$ chains. In both cases, the bulk edge has shifted to lower energy due to disorder. While the Majorana modes remain localized, in both cases there is an asymmetry in the spectral weight of the Majorana modes, reflecting the substrate disorder. Similarly, the bulk states are not periodic due to the disordered substrate.

**Discussion and outlook**

We have designed a novel type of magnet-superconducting hybrid system consisting of atom-by-atom constructed spin chains on a SC Rashba alloy with large spin-orbit coupling, thereby favoring the emergence of topological superconductivity and associated zero-energy Majorana states. Despite the presence of nanoscale LDOS variations in the BiAg$_2$ surface alloy, we clearly could observe highly localized zero-energy end states of hollow-site Fe-atom chains and finite-energy YSR bands inside the chains. In our theoretical study, we found that zero-energy end states are only expected to appear within the topologically non-trivial phase of our SC hybrid system. We took into account the effect of disorder inherent for the BiAg$_2$ surface alloy and showed that the emergent zero-energy edge states in the atomic chains are robust against chemical potential disorder, even for the case of a very small or vanishing topological gap. These results prove the robustness of zero-energy Majorana states in 1D MSH systems against disorder effects, thereby offering an explanation why Majorana modes could be observed earlier at the ends of highly disordered Fe nanowires self-assembled on SC Pb(110) substrates (*9-11*).

**ACKNOWLEDGMENTS**

**Funding:**

R.W. acknowledges financial support from the European Research Council via project No. 786020 (ERC Advanced Grant ADMIRE). J.W. and R.W. acknowledge financial support from the Deutsche Forschungsgemeinschaft (DFG) via the Hamburg Cluster of Excellence 'Advanced Imaging of Matter' (EXC 2056 - project ID 390715994). J.W. additionally acknowledges financial support from the DFG via the project WI 3097/4-1 (project No. 543483081). H.J. was supported by a postdoctoral fellowship of the Alexander von Humboldt Foundation. S.R. acknowledges support from the Australian Research Council through Grant No. DP200101118 and DP240100168. D.C. was supported by the European Union's HORIZON-RIA programme (Grant Agreement No. 101135240 JOGATE), managed by the Chips Joint Undertaking). D.C. acknowledges grants of computer capacity from the Finnish Grid and Cloud Infrastructure (persistent identifier urn:nbn:fi:research-infras-2016072533).

**Author contributions:**

H.J., L.S., J.W. and R.W. conceived and designed the experiments. H.J., L.S. and K.T.T carried out the STM/S experiments. H.J., J.W. and R.W. analyzed the experimental data. M.S. and H.O.J. performed the DFT calculations, while D.C. and S.R. performed the theoretical modeling and tight-binding calculations, and analyzed the theoretical results. J.W. and R.W. supervised the project. All authors discussed the results and contributed to the manuscript in the following way: Writing – original draft: H.J., D.C., S.R. and R.W.; Writing – review and editing: J.W., S.R. and R.W.

**Competing interests:**

The authors declare that they have no competing interests.

**Data and materials availability:**

Data sets generated during the current study are available from the corresponding authors on reasonable request.

**SUPPLEMENTARY MATERIALS**

Materials and Methods

Supplementary Text

Figs. S1 to S8

References (*31-53*)

**FIGURES**

**Figure 1.**

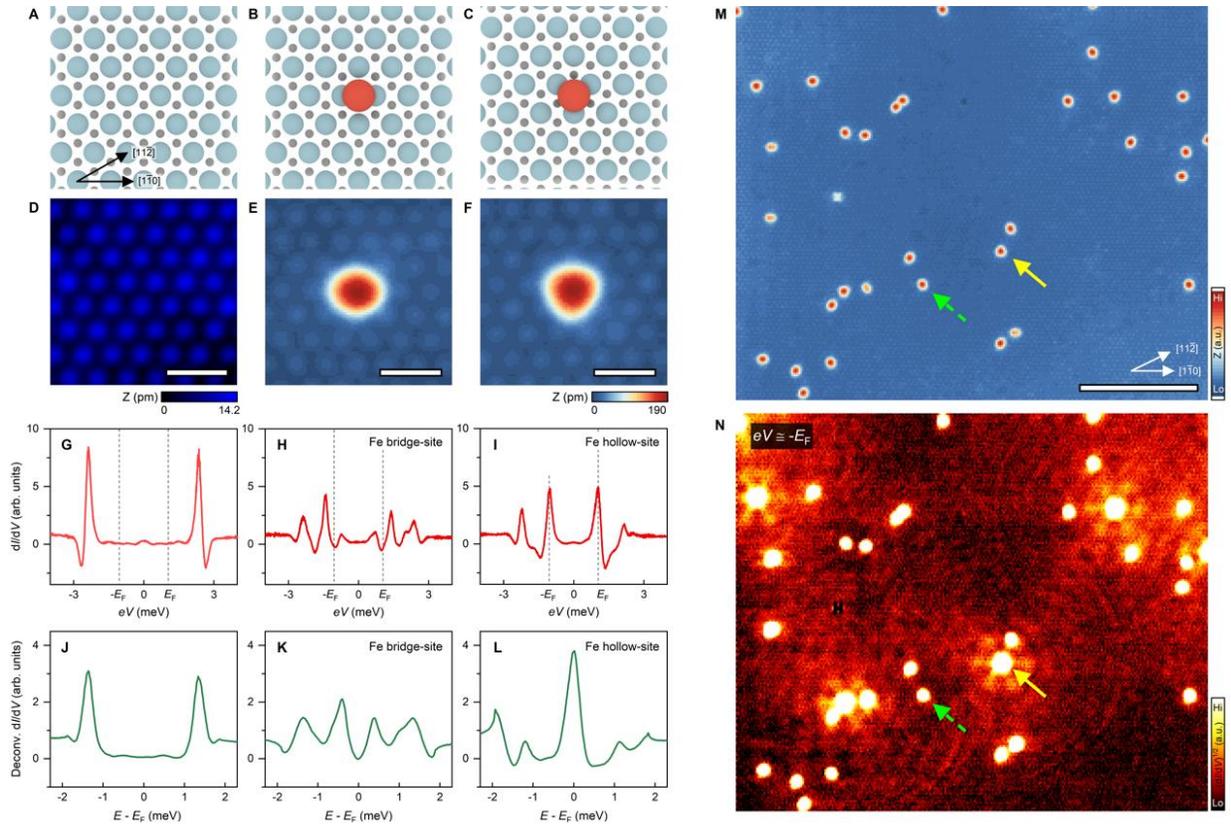

**Fig. 1. YSR states of single Fe adatoms on the BiAg$_2$ surface alloy on Ag(111)/Nb(110).** (**A** to **C**) Top views of the schematic surface structures of the ($\sqrt{3}\times\sqrt{3}$) BiAg$_2$/Ag(111)$R$30° surface alloy (**A**) without adatom and (**B**) with single Fe adatom on the bridge and (**C**) hollow site. The large red spheres at the center, the pale blue and small dark sphere symbolize Fe adatoms, as well as Bi and Ag atoms, respectively. The crystallographic directions are given in panel (**A**). (**D** to **F**) Atomically resolved constant-current STM images of (**D**) the bare BiAg$_2$ surface alloy with a Bi lattice period of 485 pm along the [1$\bar{1}$0] direction, (**E**) the Fe bridge-site adatom, and (**F**) the Fe hollow-site adatom. The white scale bars correspond to 1 nm. (**G** to **I**) Differential tunneling conductance (d$I$/d$V$) spectra as a function of energy ($eV$) measured on (**G**) the BiAg$_2$ surface alloy,

as well as on top of (**H**) the Fe bridge- and (**I**) Fe hollow-site adatoms. The vertical dashed lines correspond to the sample's Fermi level ($E_F$), which is determined based on the superconducting gap of the Nb tip. (**J** to **L**) Numerically deconvoluted d$I$/d$V$ spectra for (**J**) the BiAg$_2$ surface alloy, as well as the (**K**) Fe bridge- and (**L**) Fe hollow-site adatoms. (**M**) Constant-current STM image of the BiAg$_2$/Ag(111)/Nb(110) surface with Fe bridge-site (green dashed arrow) and Fe hollow-site (yellow arrow) adatoms. The white scale bar corresponds to 10 nm. (**N**) Spectroscopic d$I$/d$V$ map obtained for $eV = -E_F$. The YSR states of the Fe bridge-site adatoms appear highly localized on the atomic scale, while the YSR state of the Fe hollow-site adatoms exhibits long-range spatial oscillations extending over more than 10 nm. Measurement parameters: $T = 4.2$ K, $V_{mod} = 40$ $\mu$V, $V_{stab} = 5$ mV; for (**D** to **F**); $I_{stab} = 0.4$ nA, for (**G** to **I**); $I_{stab} = 1$ nA, for (**M** to **N**); $I_{stab} = 0.4$ nA.

**Figure 2.**

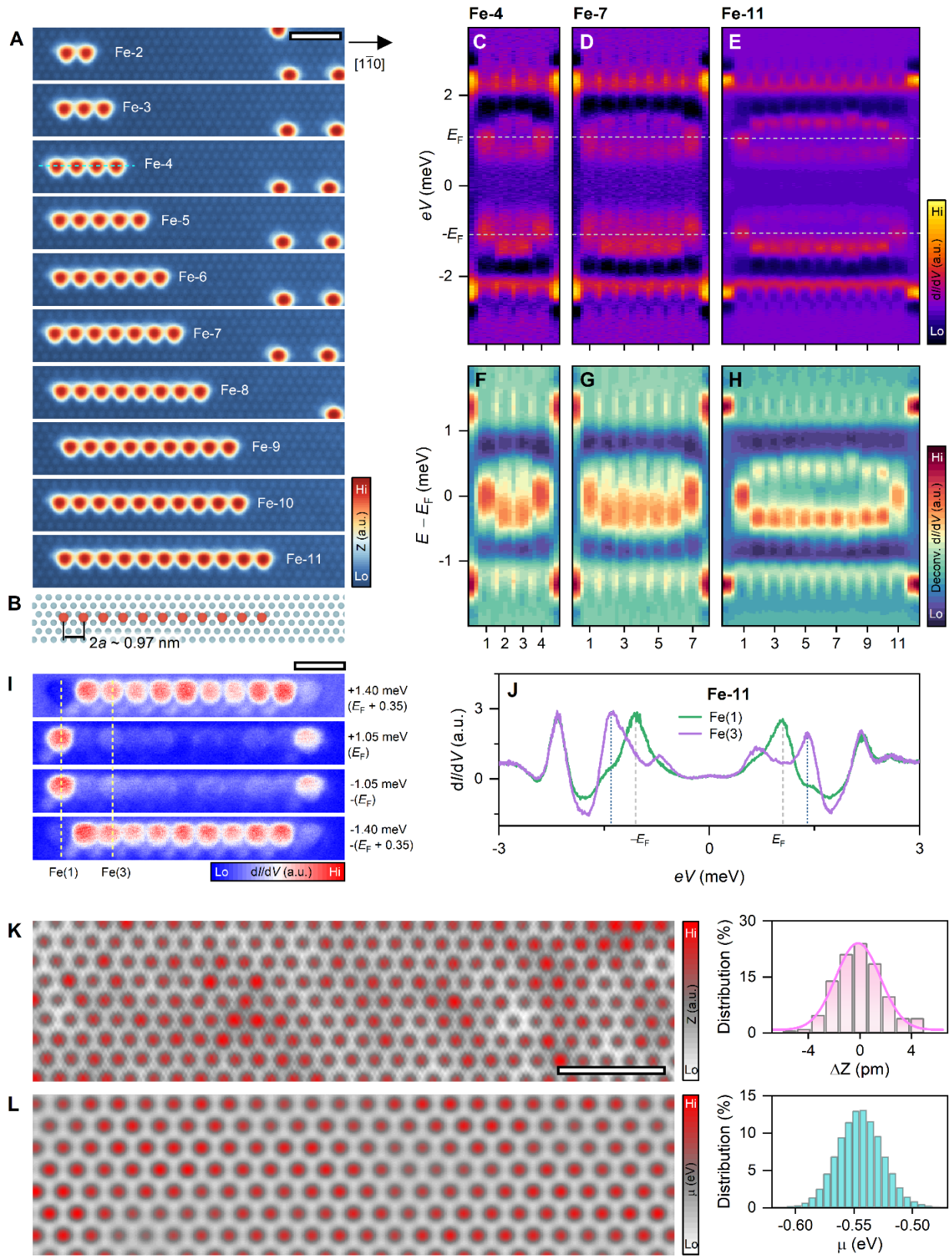

**Fig. 2. Bottom-up fabricated Fe chains on superconducting BiAg$_2$/Ag(111)/Nb(110) and their spectral characteristics.** (**A**) Constant-current STM images of artificially constructed Fe-*n* chains from Fe-2 (top) to Fe-11 (bottom) on the BiAg$_2$ surface alloy, where *n* is the number of Fe hollow-site atoms in the chain. All the distances between Fe hollow-site atoms are twice the Bi-Bi distance: $2a \sim 0.97$ nm. The white scale bar corresponds to 2 nm for panels (**A**), (**I**), and (**K**). (**B**) Schematic view of the Fe-11 chain on BiAg$_2$. (**C** to **E**) Differential tunneling conductance (d$I$/d$V$) line profiles of Fe chains along the center of the chains' axis (horizontal dashed line in (**A**) as an example) for representative Fe chains (**C**) Fe-4, (**D**) Fe-7, and (**E**) Fe-11 as-measured with a superconducting (SC) Nb-tip. The horizontal dashed lines correspond to the sample's Fermi level ($E_F$) determined from the SC gap of the Nb-tip. For panels (**C**) to (**H**), the *x*-axis denotes the sequentially numbered Fe atom from the left side of the chains. (**F** to **H**) Numerically deconvoluted d$I$/d$V$ line profiles from panels (**C**), (**D**), and (**E**) (see Suppl. Mater.). (**I**) d$I$/d$V$ maps taken on the Fe-11 chain at representative energies. The vertical dashed lines correspond to the positions of the first and third Fe atoms from the left side of the chain. (**J**) Comparison between the d$I$/d$V(V)$ spectrum for the edge Fe atom Fe(1) and the Fe(3) site. The vertical dashed and dotted lines correspond to the sample's Fermi level and $eV = \pm 1.40$ meV, respectively, i.e., the energies chosen for the d$I$/d$V$ maps in panel (**I**). (**K**) Atomically-resolved constant-current STM image of the BiAg$_2$ surface alloy on Ag(111)/Nb(110), showing nanoscale spatial variations of the local density of states. A histogram of the measured Z-height values above the Bi-atom sites is presented in the panel on the right, along with a Gaussian curve fit. (**L**) Theoretical model of the disordered BiAg$_2$ surface involving a correlated potential disorder $\mu_i$ to the substrate sites. While the tight-binding model only involves Bi sites, to aid the eye, we additionally use a cubic interpolation to plot the substrate

potential between Bi sites. A histogram of the Gaussian disorder is shown to the right. Measurement parameters: $T = 4.2$ K, $V_{stab} = 5$ mV, $V_{mod} = 40$ $\mu$V; for (**A**); $I_{stab} = 0.2$ nA, for (**C**) to (**D**); $I_{stab} = 0.5$ nA, for (**E**), (**I**), and (**J**); $I_{stab} = 1$ nA, for (**K**); $I_{stab} = 0.4$ nA. Note that a different STM tip apex was used for (**C,D**) compared to (**E**), (**I**), and (**J**), where the latter feature a slightly enhanced energy resolution.

**Figure 3.**

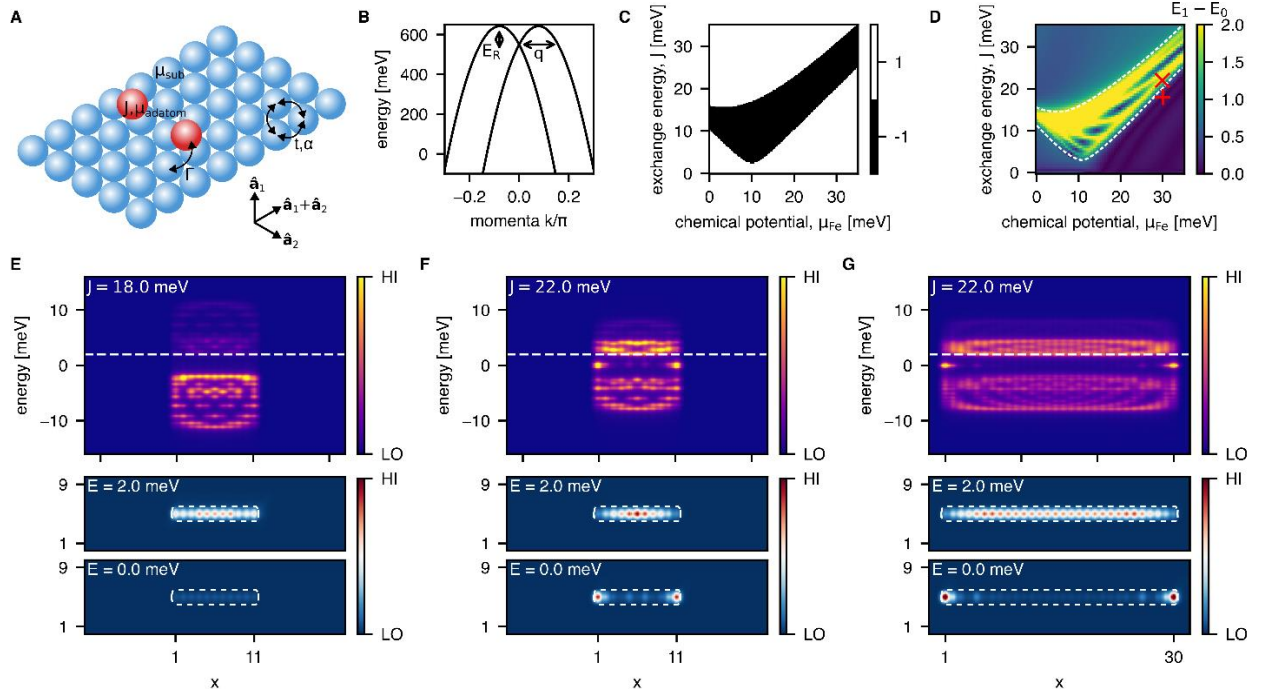

**Fig. 3. Theoretical modeling of Fe chains on superconducting BiAg₂/Ag(111)/Nb(110).** (**A**) Illustration of a triangular net formed by Bi atoms (blue), with Fe adatoms on hollow sites (red). Model parameters are indicated. Superconductivity is applied only to the Bi atoms. (**B**) Fit of the spin-split substrate bands to previous experiments on the BiAg₂ surface alloy (*28*). (**C**) Topological phase diagram for an infinite chain as a function of $\mu_{Fe}$ and $J$ ($\Gamma$ = 35 meV), with the black region being topologically non-trivial. (**D**) Topological phase diagram for a finite chain (L = 11); white dashed lines surround regions in parameter space with $E$ = 0 ground state. The color map indicates the energy gap between the ground state and the first excited state. The red cross and plus correspond to parameter pairs ($\mu_{Fe}$, $J$) for a topological and trivial regime, respectively, used for panels (E-G). (**E-G**) Local density of states (LDOS)(*E*, *x*) plots for the (**E**) trivial phase ($J$ = 18 meV, $\mu_{Fe}$ = 30 meV), (**F**) topological phase for $L$ = 11 ($J$ = 22 meV, $\mu_{Fe}$ = 30 meV), and (**G**) topological phase for $L$ = 30 ($J$ = 22 meV, $\mu_{Fe}$ = 30 meV). Bottom panels show the LDOS(*x*, *y*) plots for $E$ = 0, and for the first excited state, as indicated by the white dashed lines in the panels above.

**Figure 4.**

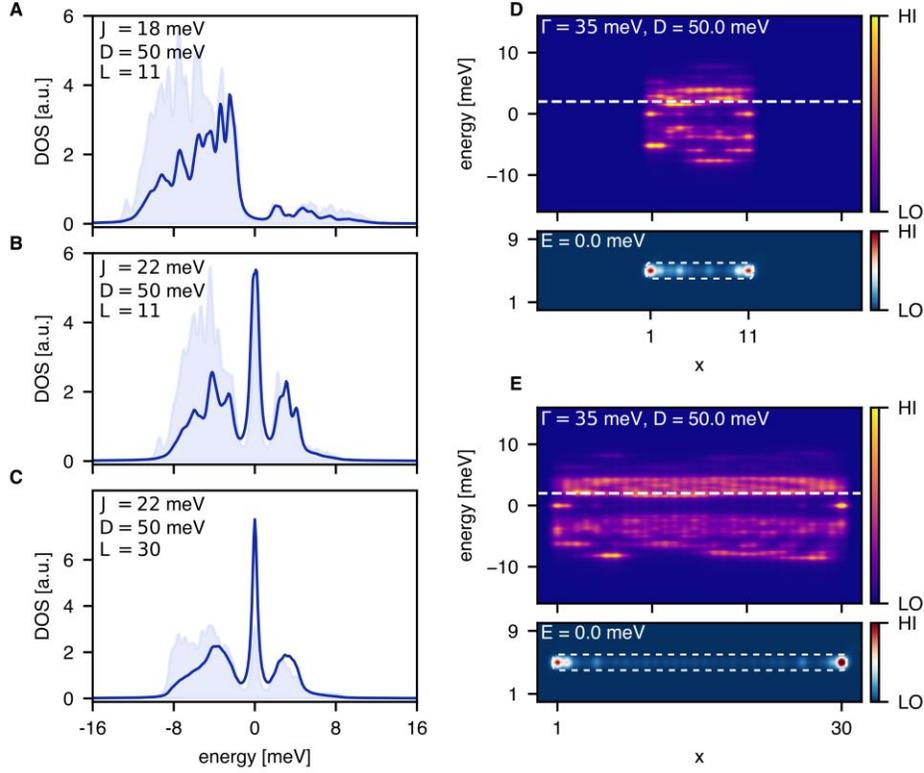

**Fig. 4. Theoretical analysis of the influence of substrate potential disorder on Majorana zero modes.** (**A** to **C**) Disorder averaged DOS at the ends of Fe chains in the (**A**) trivial phase with $L = 11$, (**B**) topological phase with $L = 11$, and (**C**) topological phase with $L = 30$. The solid lines indicate the mean and the shaded regions the standard deviation for 100 disorder realizations. (**D** to **E**) Local density of states LDOS($E, x$) plots for two disorder realizations in the topological phase for chain length $L = 11$ (**D**) and $L = 30$ (**E**). White dashed lines show the minigap edge, in the corresponding clean limit. Bottom panels show the LDOS($x, y$) plots at $E = 0$. For all plots we used $\mu_{Fe} = 30$ meV; in (D) and (E) we used $J = 22$ meV.